\documentstyle[12pt,spsfig]{article}

\def\gsim{\;
\raise0.3ex\hbox{$>$\kern-0.75em\raise-1.1ex\hbox{$\sim$}}\;
}
\def\lsim{\;
\raise0.3ex\hbox{$<$\kern-0.75em\raise-1.1ex\hbox{$\sim$}}\;
}

\catcode`\@=11
\long\def\@makefntext#1{
\protect\noindent \hbox to 3.2pt {\hskip-.9pt
$^{{\ninerm\@thefnmark}}$\hfil}#1\hfill}                %CAN BE USED

\def\@makefnmark{\hbox to 0pt{$^{\@thefnmark}$\hss}}  %ORIGINAL

\def\ps@myheadings{\let\@mkboth\@gobbletwo
\def\@oddhead{\hbox{}
\rightmark\hfil\ninerm\thepage}
\def\@oddfoot{}\def\@evenhead{\ninerm\thepage\hfil
\leftmark\hbox{}}\def\@evenfoot{}
\def\sectionmark##1{}\def\subsectionmark##1{}}

\renewcommand{\thefootnote}{\fnsymbol{footnote}}

\newcounter{sectionc}\newcounter{subsectionc}\newcounter{subsubsectionc}
\renewcommand{\section}[1] {\vspace*{0.6cm}\addtocounter{sectionc}{1}
\setcounter{subsectionc}{0}\setcounter{subsubsectionc}{0}\noindent
        {\normalsize\bf\thesectionc. #1}\par\vspace*{0.4cm}}
\renewcommand{\subsection}[1] {\vspace*{0.6cm}\addtocounter{subsectionc}{1}
        \setcounter{subsubsectionc}{0}\noindent
        {\normalsize\it\thesectionc.\thesubsectionc. #1}\par\vspace*{0.4cm}}
\renewcommand{\subsubsection}[1]
{\vspace*{0.6cm}\addtocounter{subsubsectionc}{1}
        \noindent {\normalsize\rm\thesectionc.\thesubsectionc.\thesubsubsectionc.
        #1}\par\vspace*{0.4cm}}

\newcounter{appendixc}
\newcounter{subappendixc}[appendixc]
\newcounter{subsubappendixc}[subappendixc]

\renewcommand{\appendix}[1] {\vspace*{0.6cm}
        \refstepcounter{appendixc}
        \setcounter{figure}{0}
        \setcounter{table}{0}
        \setcounter{equation}{0}
        \renewcommand{\thefigure}{\Alph{appendixc}.\arabic{figure}}
        \renewcommand{\thetable}{\Alph{appendixc}.\arabic{table}}
        \renewcommand{\theappendixc}{\Alph{appendixc}}
        \renewcommand{\theequation}{\Alph{appendixc}.\arabic{equation}}
        \noindent{\bf Appendix \theappendixc #1}\par\vspace*{0.4cm}}

\def\abstracts#1{{
        \centering{\begin{minipage}{12.2truecm}\footnotesize\baselineskip=12pt\noindent
        \centerline{\footnotesize ABSTRACT}\vspace*{0.3cm}
        \parindent=0pt #1
        \end{minipage}}\par}}

\renewenvironment{thebibliography}[1]
        {\begin{list}{\arabic{enumi}.}
        {\usecounter{enumi}\setlength{\parsep}{0pt}
\setlength{\leftmargin 1.25cm}{\rightmargin 0pt}
         \setlength{\itemsep}{0pt} \settowidth
        {\labelwidth}{#1.}\sloppy}}{\end{list}}

\newcounter{itemlistc}
\newcounter{romanlistc}
\newcounter{alphlistc}
\newcounter{arabiclistc}

\newcommand{\fcaption}[1]{
        \refstepcounter{figure}
        \setbox\@tempboxa = \hbox{\footnotesize Fig.~\thefigure. #1}
        \ifdim \wd\@tempboxa > 6in
           {\begin{center}
        \parbox{6in}{\footnotesize\baselineskip=12pt Fig.~\thefigure. #1}
            \end{center}}
        \else
             {\begin{center}
             {\footnotesize Fig.~\thefigure. #1}
              \end{center}}
        \fi}

\newcommand{\tcaption}[1]{
        \refstepcounter{table}
        \setbox\@tempboxa = \hbox{\footnotesize Table~\thetable. #1}
        \ifdim \wd\@tempboxa > 6in
           {\begin{center}
        \parbox{6in}{\footnotesize\baselineskip=12pt Table~\thetable. #1}
            \end{center}}
        \else
             {\begin{center}
             {\footnotesize Table~\thetable. #1}
              \end{center}}
        \fi}

\def\@citex[#1]#2{\if@filesw\immediate\write\@auxout
        {\string\citation{#2}}\fi
\def\@citea{}\@cite{\@for\@citeb:=#2\do
        {\@citea\def\@citea{,}\@ifundefined
        {b@\@citeb}{{\bf ?}\@warning
        {Citation `\@citeb' on page \thepage \space undefined}}
        {\csname b@\@citeb\endcsname}}}{#1}}

\newif\if@cghi
\def\cite{\@cghitrue\@ifnextchar [{\@tempswatrue
        \@citex}{\@tempswafalse\@citex[]}}
\def\citelow{\@cghifalse\@ifnextchar [{\@tempswatrue
        \@citex}{\@tempswafalse\@citex[]}}
\def\@cite#1#2{{$\null^{#1}$\if@tempswa\typeout
        {IJCGA warning: optional citation argument
        ignored: `#2'} \fi}}

 1
 1
 1

\font\ninerm=cmr9

\begin{document}

\centerline{\normalsize\bf Aperiodic Spin-Flavor Conversions and
$\tilde{\nu}_e$}
\baselineskip=16pt
\centerline{\normalsize\bf from the Sun with Random Magnetic Field}
\centerline{\scriptsize (Talk given at Ringberg EU-Conference }
\centerline{\scriptsize "New Trends in Neutrino Physics",
Ringberg, Germany, May 1998)}

%\vfill
\vspace*{0.6cm}
\centerline{\footnotesize \underline{V.B. Semikoz}~$^a$~\footnote{E-mail:
semikoz@flamenco.ific.uv.es~.  On leave from IZMIRAN, Troitsk, Moscow
region, 142092, Russia}~, A.A. Bykov~$^b$~
\footnote{E-mail: bikov@math380b.phys.msu.su}~,
 V.Yu. Popov~$^b$~\footnote{E-mail: popov@math380b.phys.msu.su},}
\baselineskip=16pt
\centerline{\footnotesize  A.I. Rez~$^c$~\footnote{E-mail:
rez@charley.izmiran.rssi.ru}~ and D.D. Sokoloff~$^b$~\footnote{E-mail:
sokoloff@lem.srcc.msu.su }}
\baselineskip=13pt
\centerline{\footnotesize\it $^a$ Instituto de F\'{\i}sica
Corpuscular - C.S.I.C.,
Departament de F\'{\i}sica Te\`orica}
\baselineskip=12pt
\centerline{\footnotesize\it Universitat de Val\`encia,
46100 Burjassot, Val\`encia, Spain}
\vspace*{0.3cm}
\centerline{\footnotesize\it $^b$ Department of Physics,
Moscow State University}
\baselineskip=12pt
\centerline{\footnotesize\it   Moscow, 119899, Russia}
\vspace*{0.3cm}
\centerline{\footnotesize\it $^c$ Institute of Terrestrial Magnetism,
Ionosphere and Radio Wave Propagation }
\baselineskip=12pt
\centerline{\footnotesize\it of the Russian Academy of Sciences, IZMIRAN}
\baselineskip=12pt
\centerline{\footnotesize\it Troitsk, Moscow region, 142092, Russia}

%\vfill
\vspace*{0.9cm}
\abstracts{The magnetic field in the solar convective zone has a random
small-scale component with the r.m.s.  value substantially exceeding
the strength of a regular large-scale field.
For two Majorana neutrino flavors $\times$ two helicities in the
presence of a neutrino transition magnetic moment and nonzero neutrino
mixing we analyze the displacement of the allowed ($\Delta
m^2$~-$\sin^2\theta$)~- parameter region reconciled for the SuperKamiokande(SK)
and radiochemical (GALLEX, SAGE, Homestake) experiments
in dependence on the r.m.s. magnetic field value $b$.
We find an effective
production of electron antineutrinos in the Sun even for small neutrino
mixing through the cascade conversions in the random magnetic field that
would be a signature of the Majorana
nature of neutrino if $\tilde{\nu}_{eR}$ will be registered.
Basing on the present SK bound on electron antineutrinos we also
estimated the corresponding excluded area in the plane
$\Delta m^2$, $\sin^22\theta$.}

%\vspace*{0.6cm}
\normalsize\baselineskip=15pt
\setcounter{footnote}{0}
\renewcommand{\thefootnote}{\alph{footnote}}
\section{Introduction and stating the problem}

Recent results of the SuperKamiokande (SK) experiment\cite{SK} have
already confirmed the solar neutrino deficit at the level less
$R_{SK}/R_{SSM}\lsim 0.4$.

Moreover, day/night (D/N) and
season neutrino flux variations were analized in this experiment and
in $\approx$ 4 years one expects to
reach
enough statistics in order to confirm or to refuse these periods
predicted in some theoretical models. D/N variations would be a signature
of the MSW solution to the Solar Neutrino Problem (SNP). The signature of the
vacuum oscillations is a seasonal variation of neutrino flux, in addition
to geometrical seasonal variation $1/r^2(t)$. And signature of the resonant
spin-flavor precession (RSFP)-solution is the 11 year periodicity of
neutrino flux.

In all four solar neutrino experiments one measures the integral spectrum,
i.e. the number of neutrino events per day for the SK experiment,
\begin{equation}
 N^{\nu}  =\sum_a \sum_i\Phi^{(0)}_i\int_{E_{th}(SK)}^{E_{max}(i)}
\lambda_i(E)\sigma_{a}(E) <P_{aa}(E)>\,dE~,
\label{integ}
\end{equation}
or the number of neutrino events  in SNU (1 SNU = $10^{-36}$ captures per
atom/per sec) in the case of GALLEX (SAGE) and Homestake experiments,
\begin{equation}
10^{-36}N^{\nu}_{Ga,~Cl} =\sum_i\Phi^{(0)}_i\int_{E_{th}(Ga,Cl)}^{E_{max}(i)}
\lambda_i(E)\sigma^{Ga,~Cl}_e (E)<P_{ee}(E)>\,dE~,
\label{radio}
\end{equation}
where $\Phi^{(0)}_i$ is the integral flux of neutrinos of  kind "i" ($i =
pp~,Be~,pep~,B$) assumed to be constant and uniform at a given distance
$r = t$ from the center of the Sun, $\lambda_i(E)$ is the  normalized
differential flux, $\int_0^{E_{max}(i)}\lambda_i(E)dE = 1$; $\sigma_{a}$
are corresponding cross sections; the thresholds $E_{th}(Ga,Cl)$ for
GALLEX (SAGE) and Homestake are $0.233~MeV$ and $0.8~MeV$ correspondingly
while for SK at present $T_{th} = 6.5~MeV$~ \cite{Bahcall}.
The probabilities
$P_{aa}(t) = \nu_a^*(t)\nu_a(t)$, where the subscript $a$ equals to
$a=e$ for $\nu_{eL}$, $a= \bar{e}$
for $\bar{\nu}_{eR}$, $a = \mu$ for $\nu_{\mu L}$ and $a=\bar{\mu}$
for $\bar{\nu}_{\mu R}$ correspondingly, satisfy the unitarity
condition
\begin{equation}
P_{ee}(t) +
P_{\bar{e}\bar{e}}(t) + P_{\mu \mu}(t) + P_{\bar{\mu}\bar{\mu}}(t) =
1~,
\label{unitarity}
\end{equation}
and~ \lq\lq$< \dots  >$\rq\rq ~corresponds to averaging over the
solar interior if necessary.

All three elementary particle physics solutions (vacuum, MSW and RSFP)
successfully describe the results of SK, GALLEX, SAGE and Homestake
solar-neutrino experiments, because the suppression factors of neutrino
fluxes are energy dependent and  helioseismic data confirm the Standard
Solar Model (SSM) with high precision at all radial distances of interest
(see recent review by Berezinsky \cite{Berezinsky}).

There exists, however, a problem with large regular solar magnetic fields,
essential for the RSFP scenario. It is commonly accepted that
magnetic fields measured at the surface of the Sun are weaker than
within interior of the convective zone where this field is supposed to be
generated. The mean field value over the solar disc is about of order
$1~G$ and in the solar spots magnetic field strength reaches $\ 1~kG$.

Because sunspots are considered to be produced from magnetic tubes
transported to the solar surface due to the boyancy, this figure can be
considered as a reasonable order-of-magnitude observational estimate for
the mean magnetic field strength  in the region of magnetic field generation.
In the solar magnetohydrodynamics \cite{Parker}
one can explain such fields in a self-consistent way if these fields are
generated by dynamo mechanism at the bottom of the convective zone (or,
more specific, in the overshoot layer). But its value
seems to be too low for effective neutrino conversions.

The mean magnetic field is however followed by a {\it small scale}, random
magnetic field. This random magnetic field is not directly traced by
sunspots or other tracers of solar activity.
This field propagates through convective zone and photosphere
drastically decreasing in the strength value with an increase of the
scale. According to the available understanding of solar dynamo, the
strength of the random magnetic field inside the convective zone is
larger than the mean field strength. A direct observational estimation
of the ratio between this strengthes is not available, however the ratio
of order 50 -- 100 does not seem impossible. At least, the ratio between
the mean magnetic field strength and the fluctuation at the solar surface
is estimated as 50 (see e.g. \cite{Stix}).

This is the main reason why we consider here an analogous to the RSFP
scenario, an aperiodic spin-flavour conversion (ASFC), based on the
presence of random magnetic fields in the solar convective zone.
It turns out that the ASFC is an additional probable way to describe
the solar neutrino deficit in different energy regions, especially if
current and future experiments will detect electron antineutrinos
from the Sun leading to conclusion that neutrinos are Majorana particles.
The termin ``aperiodic'' simply reflects the exponential
behaviour of conversion probabilities in noisy media (cf. \cite{Nicolaidis}
, \cite{Enqvist}).

As well as for the RSFP mechanism all arguments for
and against the ASFC mechanism with random magnetic fields remain the same
ones that have been recently summarized and commented by Akhmedov
(see \cite{Akhmedov1} and references therein).

But contrary to the case of regular magnetic fields we find out that
one of the signatures of random magnetic field in the Sun is the
prediction of a wider allowed region for neutrino mixing angle in the
presence of $\bar{\nu}_e$ from the Sun including the case of {\it
small mixing angles}.

\section{Magnetic Field Model and Numerical Simulation}

The random magnetic field is considered to be
maximal somewhere at the bottom of convective zone and decaying to the
solar surface. To take into account a possibility, that the solar dynamo
action is possible also just below the bottom of the convective zone
(see \cite{Parker}),
we accept, rather arbitrary, that it is distributed at
the radial range $0.7 R_\odot~ \mbox{---}~  1.0 R_\odot$, i.e. it has the same
thickness as the convective zone, while its correlation length is
$L_0=10^4~km$ that is close to the mesogranule size.

We suppose also that
all the volume of the convective zone is covered by a net
of rectangular domains where the random magnetic field
strength vector is constant. The magnetic field strength changes
smoothly at the boundaries between the neighbour domains obeying the
Maxwell equations. Since one can not expect the strong influence of
small details in the random magnetic field within and near thin boundary
layers between the domains this oversimplified model looks applicable.

In agreement with the SSM \cite{Bahcall}
the neutrino source is supposed  to be located inside

\psfig{file=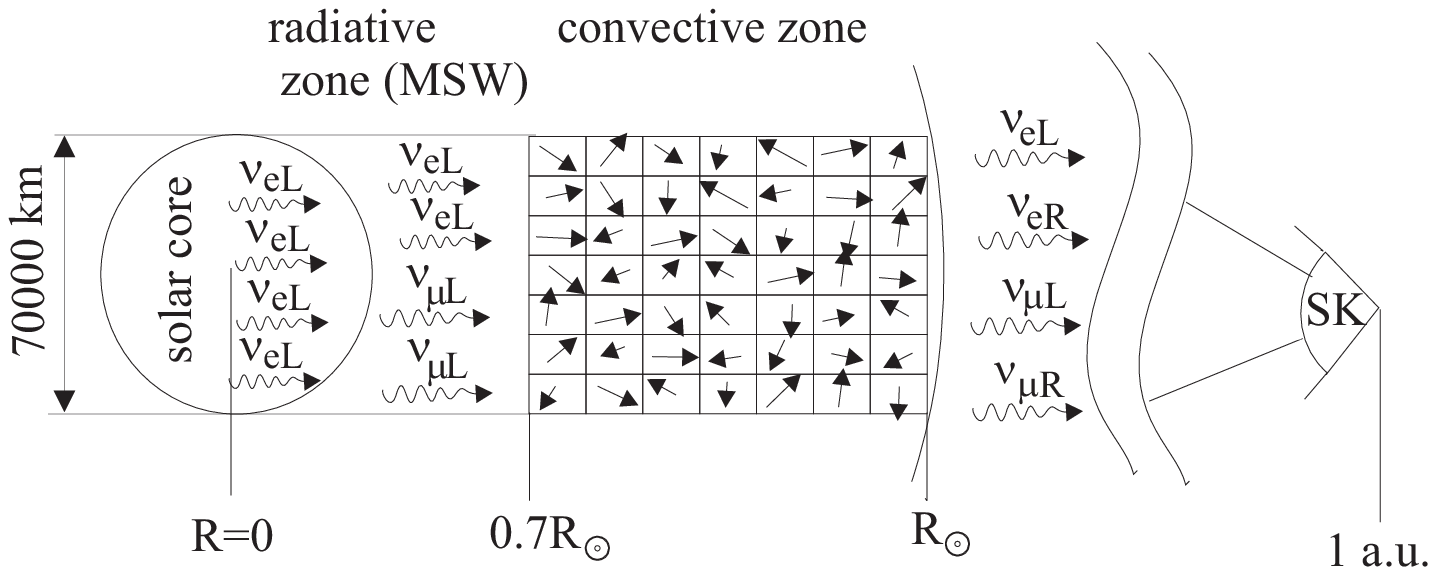}
\begin{quote}
{\small
Fig 1. Geometry of neutrino trajectories in random magnetic fields.
Finite radius of $\nu_{eL}$-neutrino source, the solar core radius
$0.1R_{\odot}$, is shown.}
\end{quote}

\noindent the solar
core with the radius of the order of $R_{\nu} = 0.1R_{\odot}= 7\times
10^4~km$. For the sake of simplicity we do not consider here
the spatial distribution of neutrino emissivity from a unit of a solid
angle of the  core image and the differences in $R_{\nu}^{(i)}$ for
different neutrino kinds "i".
Different parallel trajectories directed to the
Earth cross different magnetic domains because the domain size $L_0$
(in the plane which is perpendicular to neutrino trajectories) is
much
less than the transversal size $=R_{\nu}$ of the full set of parallel
rays, $L_0\ll R_{\nu}$, see Fig. 1.
The whole number of  trajectories (rays) with statisticaly
independent magnetic fields is about $R_{\nu}^2/L_0^2\sim 50$.

At the stage of numerical simulation of the random magnetic field
we generate the set of random numbers with a given r.m.s. value
and for each realization of random
magnetic fields along each of $50$ rays solve the Cauchi
problem for the $(4 \times 4)$ master equation ,

\begin{equation}
i\left(
\begin{array}{l}
\dot{\nu}_{eL}\\
\dot{\tilde{\nu}}_{eR}\\
\dot{\nu}_{\mu L}\\
\dot{\tilde{\nu}}_{\mu R}
\end{array}
\right) =
\left(
\begin{array}{cccc}
V_e -c_2\delta & 0 & s_2\delta & \mu H_+(t) \\
0 & - V_e - c_2\delta & - \mu H_-(t)& s_2\delta \\
s_2\delta & - \mu H_+(t)&V_{\mu} + c_2\delta &0 \\
\mu H_-(t)&s_2\delta &0& - V_{\mu} + c_2\delta
\end{array}
\right)
\left(
\begin{array}{c}
\nu_{eL}\\
\tilde{\nu}_{eR}\\
\nu_{\mu L} \\
\tilde{\nu}_{\mu R}
\end{array}
\right)~,
\label{master}
\end{equation}
where $c_2 = \cos 2\theta$, $s_2 = \sin 2\theta$, $\delta = \Delta
m^2/4E$ are the neutrino mixing parameters; $\mu = \mu_{12}$ is the
neutrino active-active transition magnetic moment;
$H_{\pm}(t) = B_{0\pm}(t) + b_{\pm}(t)$, $H_{\pm} = H_x \pm
iH_y$, are the magnetic field components consisting of regular
($B_{0\pm}$) and random ($b_{\pm}$) parts which are
perpendicular to the neutrino trajectory in the Sun; $V_e(t) =
G_F\sqrt{2}(\rho (t)/m_p)(Y_e - Y_n/2)$ and $V_{\mu}(t) =
G_F\sqrt{2}(\rho (t)/m_p)(- Y_n/2)$ are the neutrino vector
potentials
for $\nu_{eL}$ and $\nu_{\mu L}$ in the Sun given by the abundances
of
the electron ($Y_e = m_pN_e(t)/\rho (t)$) and neutron ($Y_n =
m_pN_n(t)/\rho (t)$) components and by the SSM density profile $\rho
(t) = 250~g/cm^{3~}\exp ( - 10.54t)$ \cite{Bahcall}.
Then we  calculate the dynamical probabilities $P_{aa}(t) $, and average
them in transversal plane, i.e. obtain  the mean arithmetic probabilities
as functions of  mixing parameters $\sin^22\theta$ , $\delta $. We argue
that being additive functions of the area of
the convective zone layer (or, the same, of the number of rays) these
probabilities with increasing area become self-averaged.

%%%%%%%%%%%%%%%%%%%%%%%%%%%%%%%%%%%%%%%%%%%%%%%%%%%%%%%%%%%%%%%%%%%
\section{Asymptotic solution }

To demonstrate the properties of our model we briefly discuss
how random magnetic field influences the small-mixing
MSW solution to SNP.
It turns out that for the SSM exponential density  profile,
typical borone neutrino energies
$E \sim 7\div14 MeV$,
and
$\triangle m^2\sim10^{-5}~eV^2$,
the MSW resonance
(i.e. the point where $V_e-V_{\mu}=2c_2\delta)$
occurs well below the bottom of the convective zone.
Thus  we  can  divide  the  neutrino
propagation problem and  consider  two  successive  stages.
First, after generation in the middle of the Sun,
neutrinos propagate in the absence of any magnetic  fields,
undergo  the non-adiabatic (non-complete) MSW  conversion
$\nu_{eL} \to \nu_{\mu L}$
and  acquire  certain  nonzero   values
$P_{ee}$  end  $P_{\mu\mu}$,
which  can  be  treated  as   initial conditions
at the bottom of the convective zone.
For small  neutrino mixing
$s_2\to 0$  the  $(4\times4)$-master  equation  Eq.( \ref{master}) then splits
into  two  pairs  of  independent   equations
describing correspondingly  the   spin-flavor    dynamics
$\nu_{eL} \to \tilde\nu_{\mu R}$
and
$\nu_{\mu L} \to  \tilde\nu_{eR}$
in  noisy magnetic fields.
In addition, once the MSW resonance point is far away
from the convective zone,
one can also omit $V_e$ and $V_{\mu}$ in comparison with $c_2\delta$.
For $\nu_{eL}\to \tilde\nu_{\mu R}$ conversion
this results into a  two-component  Schr$\rm \ddot o$dinger equation
\begin{equation}
  i\left(\begin{array}{c}
    \dot\nu_{eL}\\
    \dot{\tilde\nu}_{\mu R}
         \end{array}\right)
  \simeq \left(\begin{array}{cc}
   -\delta & \mu b_{+}(t)\\
    \mu b_{-}(t) & \delta
               \end{array}\right)
  \left(\begin{array}{c}
    \nu_{eL}\\
    \tilde\nu_{\mu R}
        \end{array}\right)
\label{(01)}
\end{equation}
with initial conditions
$
  |\nu_{eL}|^2(0)=P_{ee}(0),
  \quad
  |\tilde\nu_{\mu R}|^2 = P_{\tilde\mu\tilde\mu}(0) =0.
$
As normalized probabilities
$P_{ee}(t)$
and
$P_{\tilde\mu\tilde\mu}(t)$
(satisfying the conservation law
$P_{ee}(t) + P_{\tilde\mu \tilde\mu}(t) =  P_{ee}(0)$)
are the only observables,
it is convenient to recast Eq.(\ref{(01)})
into an equivalent integral form
\begin{equation}
\begin{array}{rcl}
  \displaystyle
  S(t) &=& S(0) - 4\mu^2
    \int \limits_0^t dt_1
    \int \limits_0^{t_{1}} dt_2
    S(t_2)
    \\
  \displaystyle
    &&\{
     [b_x(t_1)b_x(t_2) + b_y(t_1)b_y(t_2)] \cos2\delta(t_1 - t_2)
  \displaystyle
    \\
    &&-[b_x(t_1)b_y(t_2) - b_x(t_2)b_y(t_1)] \sin2\delta(t_1 - t_2)
    \},
\end{array}
\label{(03)}
\end{equation}
\noindent
where
$S(t)=2P_{ee}(t)-P_{ee}(0) \equiv P_{ee}(t)-P_{\tilde\mu\tilde\mu}(t)$
is the third component of the polarization vector $\vec S$.

Let us trace the line of further derivation in a spirit of our
numerical method.
Dividing the interval of integration
into a set of  equal intervals
of correlation length $L_0,$
and assuming  that possible correlations between
$S(t_2)$ and $b_i , b_j$ under the integral are  small,
$S(t)$
itself varies very slowly within one correlation cell,
and making  use  of    statistical properties
of random fields,
($i$) different  transversal  components
within one  cell  are  independent  random  variables,  and
($ii$)
magnetic fields in different cells do not correlate,
we can average Eq.(\ref{(03)})
thus obtaining a finite difference analogue.
Returning back to continuous version we get
\begin{equation}
  S(t) = S(0) \exp \left\{ -\frac43 \mu^2L_0
    \frac{\sin^2\delta  L_0}{(\delta  L_0)^2}
    \int\limits^t_0 <\vec b^2(t')> dt'\right\}~,
\label{finite}
\end{equation}
where we retained possible slow space  dependence
of the  r.m.s. magnetic field value.

For $\delta~ \ll L_0^{-1}$
and  constant~  r.m.s.~  $<\vec b^2>$~
we  obtain  the~  simple~  $\delta$-correlation~  result
 \cite{Nicolaidis} , \cite{Enqvist}, otherwise there remains
an additional stabilizing factor $\sin^2x/x^2$, demonstrating
to what extent the vacuum/medium neutrino oscillations within
one correlation cell can suppress the spin-flavor dynamics due
to the magnetic field  only.

Another important issue
is the problem of temporal dependence of higher statistical moments
of $P_{aa}.$
As $P_{aa}$ enter the Eq.(\ref{integ}) for the number of events
one should be certain that the averaging procedure
does not input large statistical errors,
otherwise there will be no room
for the solar neutrino puzzle itself.
For $\delta \ll L_0^{-1}$ basing on Eq. ( \ref{(03)}) we evaluate
the dispersion $\sigma^2_P$. Here we present only the final result,
the details of derivation will be published elsewhere:
\begin{equation}
  \sigma_P = \sqrt{<P_{ee}^2(t)> - <P_{ee}(t)>^2}
  = \frac{P_{ee}(0)}{2\sqrt2} \left( 1 -  e^{- 2 \Gamma t} \right)~,
  \label{(C8)}
\end{equation}

\begin{equation}
  <P_{ee}(t)> = \frac{P_{ee}(0)}{2} \left( 1 + e^{- \Gamma t} \right),
  \label{(C9)}
\end{equation}
where  $ \Gamma =(4/3) \mu^2 <\vec b^2>L_0 $.
We see that relative mean square deviation of $P_{ee}$
from its mean value tends with $t\to\infty$
to its maximum asymptotic value
\begin{equation}
  \frac{\sigma_P(t)}{<P_{ee}(t)>}
  \to \frac{1}{\sqrt2} \simeq 0.707 , \qquad t \to \infty,
  \label{(C10)}
\end{equation}
irrespectively of the initial value $P_{ee}(0).$

This estimate is true, evidently, only for one neutrino ray.
Averaging over $N$ independent rays lowers the value (\ref{(C10)})
in $\sqrt N$ times.
That is for our case of $N \approx 50$ rays we get that
maximum relative error should not exceed approximately $10\%,$
thus justifying the validity of our approach.
For smaller magnetic fields the situation is always better.

To conclude this section, it is neccesary to repeat that the above
estimate Eq. (\ref{(C10)}) indicates possible danger when treating
numerically the neutrino propagation in noisy media. Indeed,
usually adopted one-dimensional (i.e. along one ray only)
approximation for the $(4 \times 4)$ master equation Eq. (\ref{master})
or $(2 \times 2)$ Eq. (\ref{(01)}) can suffer from large dispersion
errors and one should make certain precautions when averaging these
equations over the random noise {\it before} numerical simulations.
Otherwise, the resulting error might be even unpredictable.

\section {Discussion of results}

In order to find the regions in the $\Delta m^2$, $\sin^22\theta$
plane excluded from nonobservation of $\bar{\nu}_{eR}$ in SK
it is necessary to compute and plot the isolines of the ratio
$\Phi_{\bar{\nu}}/\Phi_B^{SSM}$.
Indeed, as antineutrinos were not detected in the SK experiment one should
claim that antineutrino flux is smaller than the SK background
$\Phi_{\bar{\nu}_e}(E>8.3~MeV) < 6 \times 10^4~cm^{-2}s^{-1}$~
\cite{Fiorentini}. Deviding this inequality by the SSM borone
neutrino flux with energies $E>E_{th}=8.3~MeV$,~~
$\Phi_B^{SSM}(E>8.3~MeV) = 1.7\times 10^6~cm^{-2}s^{-1}$ we find
the bound on the averaged over cross-section and spectrum, cascade
transition probability $\nu_{eL}\to \bar{\nu}_{eR}$~,
\begin{equation}
\frac{\Phi_{\bar{\nu}_e}(E>8.3~MeV)}{\Phi_B^{SSM}(E>8.3)} =
\frac{\int_{8.3}^{15}\lambda_B(E)P_{\bar{e}\bar{e}}(E)
\sigma_{\bar{\nu}}(E)dE}{\int_{8.3}^{15}\lambda_B(E)
\sigma_{\bar{\nu}}(E)dE}\lsim 0.035~,
\label{percent}
\end{equation}

\begin{tabular}{p{7.4cm}p{7.4cm}}
\vskip 0.7cm
\psfig{file=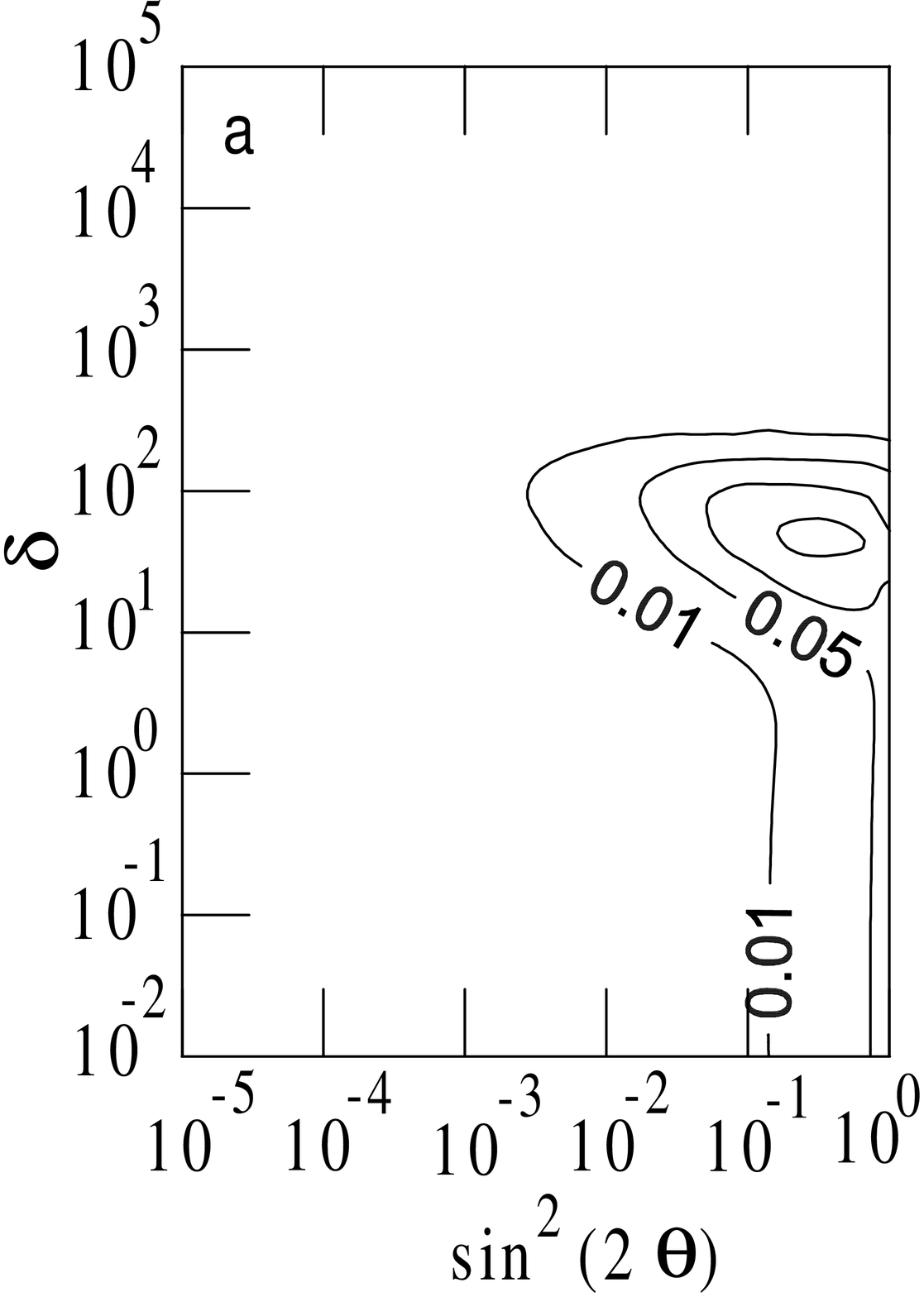,height=8.5cm,width=6.1cm} &
\vskip 0.7cm
\psfig{file=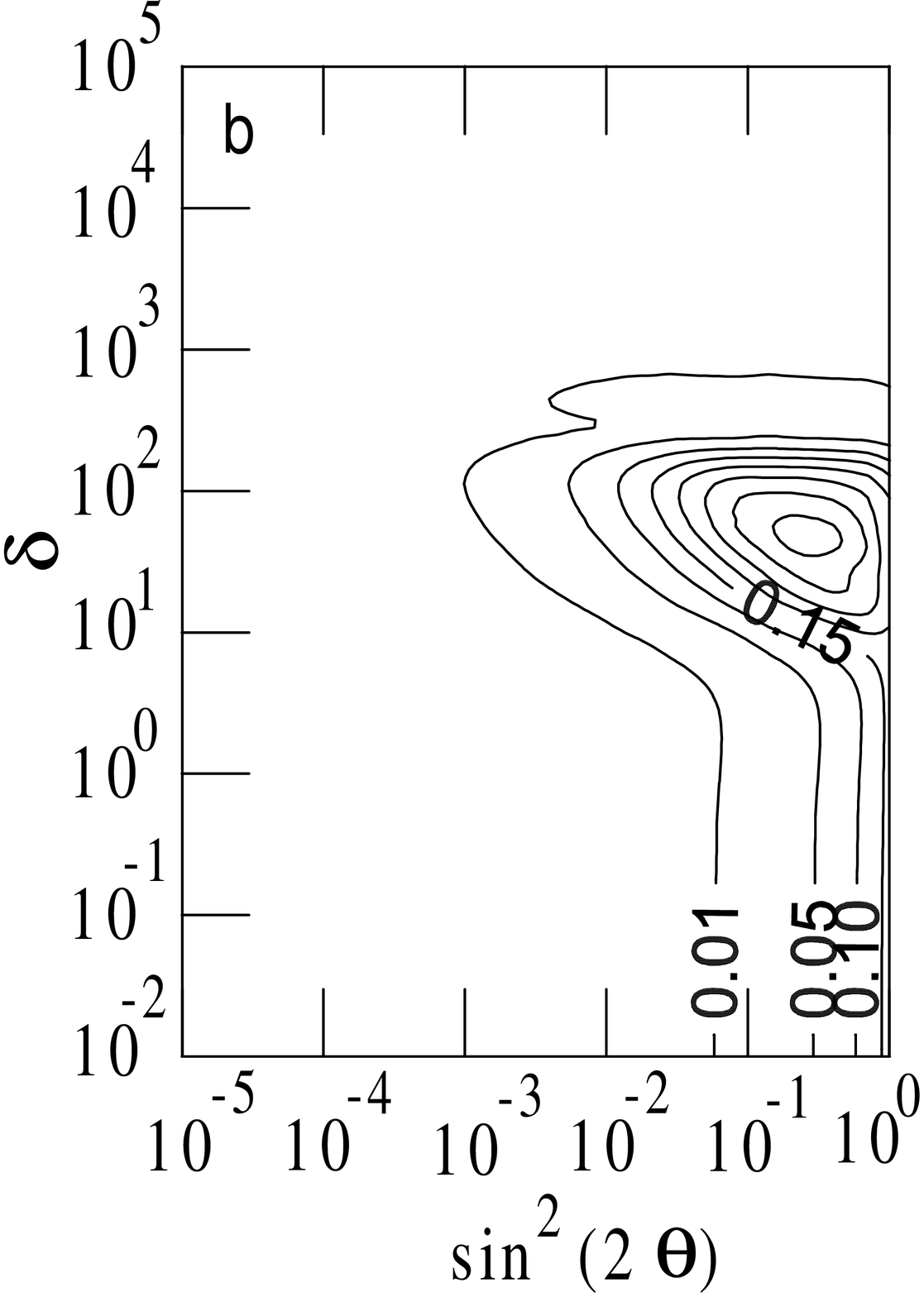,height=8.5cm,width=6.1cm} \\
\centerline{(a)} & \centerline{(b)}
\end{tabular}
\begin{quote}
{\small
Fig 2. The averaged (over 50 rays) probability
$P_{\bar{e}\bar{e}} = \langle \bar{\nu}_{eR}^*\bar{\nu}_{eR}\rangle$
in dependence on $\delta = \Delta m^2/4E$ and
$\sin^22\theta$-parameters for $b_{r.m.s.}= 50~kG$ (a) and $b_{r.m.s.}=
100~kG$ (b). The dimensionless $\delta =
8.81\times 10^8\Delta m^2(eV^2)/E(MeV)$ is plotted along y-axis.} \end{quote}
\vskip 0.3cm
or $\Phi_{\bar{\nu}}/\Phi_B^{SSM}\lsim 3.5$\%~.
Here $\sigma_{\bar{\nu}}(E) = 9.2\times 10^{-42}~cm^2[(E -
1.3~MeV)/10~MeV]^2$ is the cross-section of the capture reaction
$\bar{\nu}_ep\to ne^+$.
 Notice that the above bound is
not valid for low energy region below
the threshold of the antineutrino capture by protons, $E\leq 1.8~MeV$.

For low magnetic fields, $B~,b\lsim 20~kG$, both for regular
and random ones we do not find any violation of the bound Eq.
(\ref{percent}).
However, for strong magnetic fields such forbidden parameter regions
appear in different areas over $\delta $ and
$\sin^22\theta$ for different kind of magnetic fields (regular and random).
In general, the influence of random magnetic fields is more pronounced
as compare to the regular ones of the same strength. Here we illustrate
only the random magnetic field case, see Fig. 2,3 for $b_{rms} = 50~
\mbox {and}~100~kG$.
It follows that the more intensive  the r.m.s. field
$\sqrt{\langle b^2\rangle}$ in the convective zone the more effective
spin-flavour conversions lead to the production of the right-handed
$\bar{\nu}_{eR}$, $\bar{\nu}_{\mu R}$-antineutrinos. In the case of small
mixing this is also  seen from Eq. (\ref{finite}).

\begin{tabular}{p{7.4cm}p{7.4cm}}
\vskip 0.7cm
\psfig{file=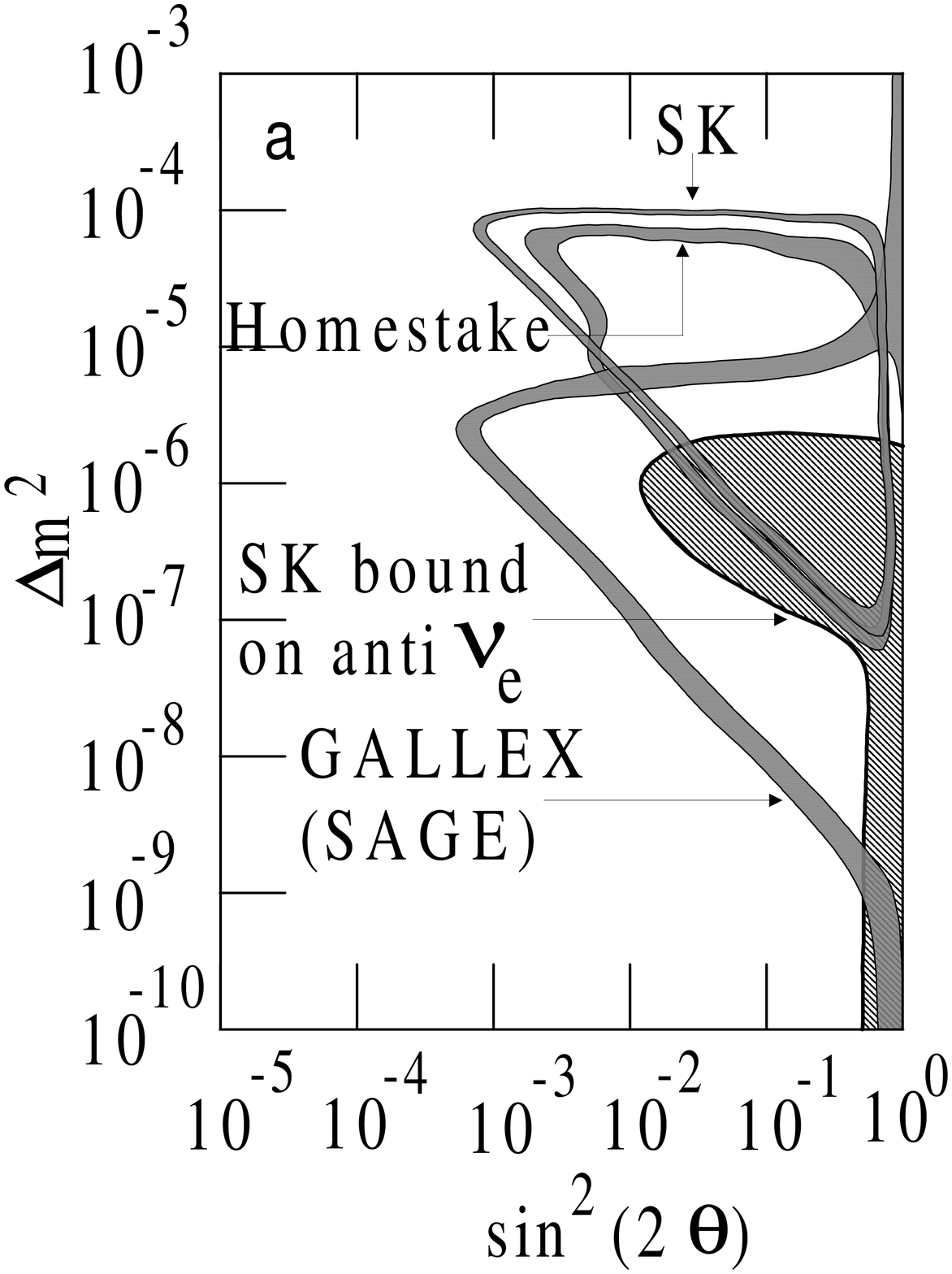,height=8.5cm,width=6.1cm} &
\vskip 0.7cm
\psfig{file=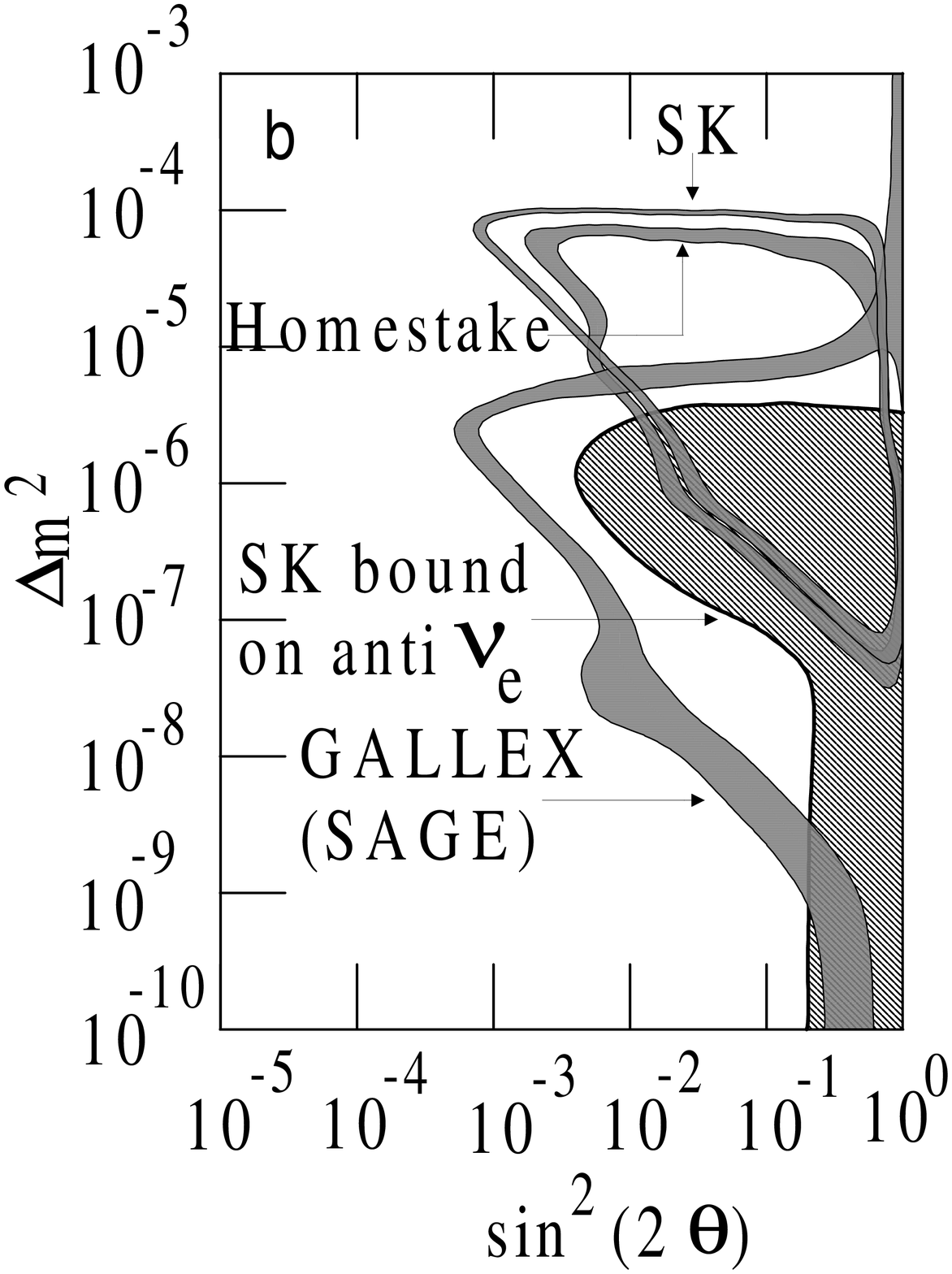,height=8.5cm,width=6.1cm} \\
\centerline{(a)} & \centerline{(b)}
\end{tabular}
\begin{quote}
{\small
Fig 3. The reconcilement of all experiments in the plane $\Delta
m^2(eV^2)$ and $\sin^22\theta$.  The isolines with dashed bands represent
the DATA/SSM ratio equal correspondingly to $0.368\pm 0.026~$(SK),
$0.509\pm 0.059$, $0.504 \pm 0.085$ (GALLEX~ + SAGE),~and $0.274\pm
0.027~$(Homestake).\\ (a) The random field $b_{rms}= 50~kG$; (b) the
random field $b_{rms}= 100~kG$.} \end{quote} \vskip 0.3cm In Fig. 3 we
present the allowed parameter regions to reconcile four solar neutrino
experiments along with the SK bound on $\bar{\nu}_{eR}$ given by Eq.
(\ref{percent}). One can see that without the latter bound there exist two
commonly adopted parameter regions, the small mixing and the large mixing
ones, currently allowed or excluded in dependence of the values of the
r.m.s. magnetic field parameters. Here we demonstrate the allowed case
only, when excluded by the SK bound region is well below the two parameter
regions of interest.

Our results, however, show that there exists strong
dependence not only on the r.m.s. magnetic field strength but on its
correlation length also. In particular, when for the same $b_{rms} =
100~kG$ we substitute $L_0 =10^3~km$ (a granule size) instead of $L_0 =
10^4~km$ (a mesogranule size treated in numerical simulation) the small
mixing MSW region is fully excluded via the SK bound Eq. (\ref{percent}).
This is easily explained due to the maximum value of ASFC seen from Eq.
(\ref{finite}) ($\sin x/x\to 1$).  Vice versa, if the small-mixing MSW
solution is valid and a large neutrino magnetic moment exists we can
extract from neutrino data important constraints on the structure of
random magnetic fields deep under surface of the Sun.\vskip 0.3cm
%%%%%%%%%%%%%%%%%%%%%%%%%%%%%%%%%%%%%%%%%%%%%%%%%%%%%%%%%%%%%%%%%%%%%%%%
\section{Conclusions}

We develop a model of neutrino spin-flavour conversions in the
random magnetic fields of the solar convective zone supposing
in consistence with modern MHD models of solar magnetic fields
that random fields are naturally much higher than large-scale magnetic
fields~ created and supported~ continuously from the small-scale~ random ones
\cite{Parker}.

It follows that if neutrinos have a large transition magnetic moment
their dynamics in the Sun is governed by random
magnetic fields that , first, lead to {\it aperiodic} and rather
{\it non-resonant} neutrino spin-flavor conversions, and second, inevitably
lead to production of electron antineutrinos for {\it low energy} or
{\it large mass difference} region.
If antineutrinos $\bar{\nu}_{eR}$ would be found with the positive
signal in the BOREXINO experiment\cite{BOREXINO} or, in other words,
a small-mixing MSW solution to SNP fails, this will be a strong argument
in favour of magnetic field scenario with ASFC in the presence of a
large neutrino transition moment, $\mu\sim 10^{-11}\mu_B$ for the same
small mixing angle.

The search of  bounds on $\mu$ at the level $\mu\sim 10^{-11}\mu_B$
in low energy $\nu~e$-scattering, currently planning in laboratory
experiments \cite{LAMA}, will be crucial for the model considered here.

We would like to emphasize the importance of future low-energy neutrino
experiments (BOREXINO, HELLAZ) which will be sensitive both to check
the MSW scenario and the $\bar{\nu}_{eR}$-production through ASFP.
As it was shown in a recent work\cite{PSV} a different slope of energy
spectrum profiles for different scenarios would be a crucial test in
favour of the very mechanism providing the solution to SNP.
\vskip 0.3cm
\centerline{\bf Acknowledgements}
\vskip 0.3cm
The authors thank Sergio Pastor, Emilio Torrente, Jose Valle and Klaus
Winter for fruitful discussions. This work has been supported by RFBR
grant 97-02-16501 and by INTAS grant 96-0659 of the European Union.

\vskip 0.3cm

%%%%%%%%%%%%%%%%%%%%%%%%%%%%%%%%%%%%%%%%%%%%

\end{document}